\documentclass[aps,prl,floatfix,twocolumn]{revtex4}

\usepackage{amsmath}
\usepackage{amssymb}
\usepackage{graphicx}
\usepackage{epstopdf}
\usepackage{psfrag}
\usepackage[usenames]{color}
\usepackage{hyperref}
\usepackage{bm}
\usepackage{bbm}
\usepackage{soul}

\newcommand{\ket}[1]{\vert#1\rangle}

\newcommand{\sub}[1]{_{\rm #1}}

\begin{document}
\title{Implementation of 
a non-deterministic optical noiseless amplifier}
\author{Franck Ferreyrol,  Marco Barbieri, R\'emi Blandino, Simon Fossier, Rosa Tualle-Brouri, and Philippe Grangier.}
\affiliation{Groupe d'Optique Quantique, Laboratoire Charles Fabry, Institut d'Optique, CNRS, Universit\'e Paris-Sud, Campus Polytechnique, RD 128, 91127 Palaiseau cedex, France }
\begin{abstract}
Quantum mechanics imposes that any amplifier that works independently on the phase of the input signal has to introduce some excess noise. The impossibility of such a noiseless amplifier is rooted into unitarity and 
linearity of quantum evolution. A possible way to circumvent this limitation is to interrupt such evolution via a measurement, providing a random outcome able to ÔheraldÕ a successful - and noiseless - amplification event. Here we show a successful realisation of such an approach; we perform a full characterization of an amplified coherent state using quantum homodyne tomography, and observe a strong ÔheraldedÕ amplification, with about 6dB gain and a noise level significantly smaller than the minimal allowed for any ordinary phase-independent device. 
\end{abstract}
\maketitle
Quantum optical detection techniques are so advanced that quantum fluctuations are the main source of noise. Therefore, when amplifying optical signals, one has to look at intrinsic limitations of the process: any amplifier cannot work independently on the phase of the input, unless some additional noise is added \cite{caves82}. The origin of this limitation is that adding extra noise is needed for the output field to obey Heisenberg's uncertainty relation. Also, it is connected to the impossibility of realizing arbitrarily faithful copies of a quantum signal \cite{wootters82, and05}, and it is thus deeply rooted in the linear and unitary evolution of quantum mechanical systems.

Various aspects of this limitation have been studied by using optical parametric amplifiers\cite{levenson93,levenson93a,braunstein01,cochrane04}. For instance, a non-degenerate optical parametric amplifier amplifies all input phases, and introduces the minimal level of added noise, which degrades the signal-to-noise ratio \cite{caves82}. The same process, driven in the degenerate regime, may provide amplification preserving the signal-to-noise ratio. However, this occurs in a phase-dependent fashion: only the part of the signal in phase with the pump light will be amplified, while the part which is 90 degrees out of phase with the pump will be de-amplified\cite{levenson93,levenson93a}.

A more intriguing idea is to find a way to tamper with the linear evolution of quantum mechanics; this is actually possible, though non-deterministically, by conditioning our observation upon the result of a measurement\cite{Knill01}. Noiseless amplification can then take place, but only a fraction of the times, and the ÔcorrectÕ operation is heralded. This strategy is commonly adopted for building effective nonlinearities in linear quantum optical gates\cite{obrien03, lanyon09}.

Here we follow the proposal of Ralph and Lund \cite{ralph09} to demonstrate experimentally that heralded non-deterministic amplification  
 can realise processes which would 
 be impossible for usual amplifiers. Unlike another realisation \cite{xiang09}, we have direct access to the output state via state tomography, so we can provide a complete description of the process, and analyse the limitations arising from non-ideal components. Our study is relevant in the long-term view of the integration of amplifiers in quantum communication lines \cite{fossier09}.

The conceptual layout of the noiseless amplifier is presented in Fig. 1. The operating principle is closely related to quantum teleportation\cite{bennett93,bouw97,boschi98,furusawa98}, and is actually a variation of the quantum scissors protocol \cite{pegg98,babichev03}: the phase and amplitude information of the input are transferred via a generalised teleportation onto a superposition of the vacuum and a single photon. If the input is not too large, such superposition is still adequate to describe a coherent state with a good fidelity. The amplification is allowed by the use of a non-maximally entangled resource \cite{ralph09}.

\begin{figure}[b]
\includegraphics[viewport = 80 300 600 680, clip, width=.7\columnwidth, angle=0]{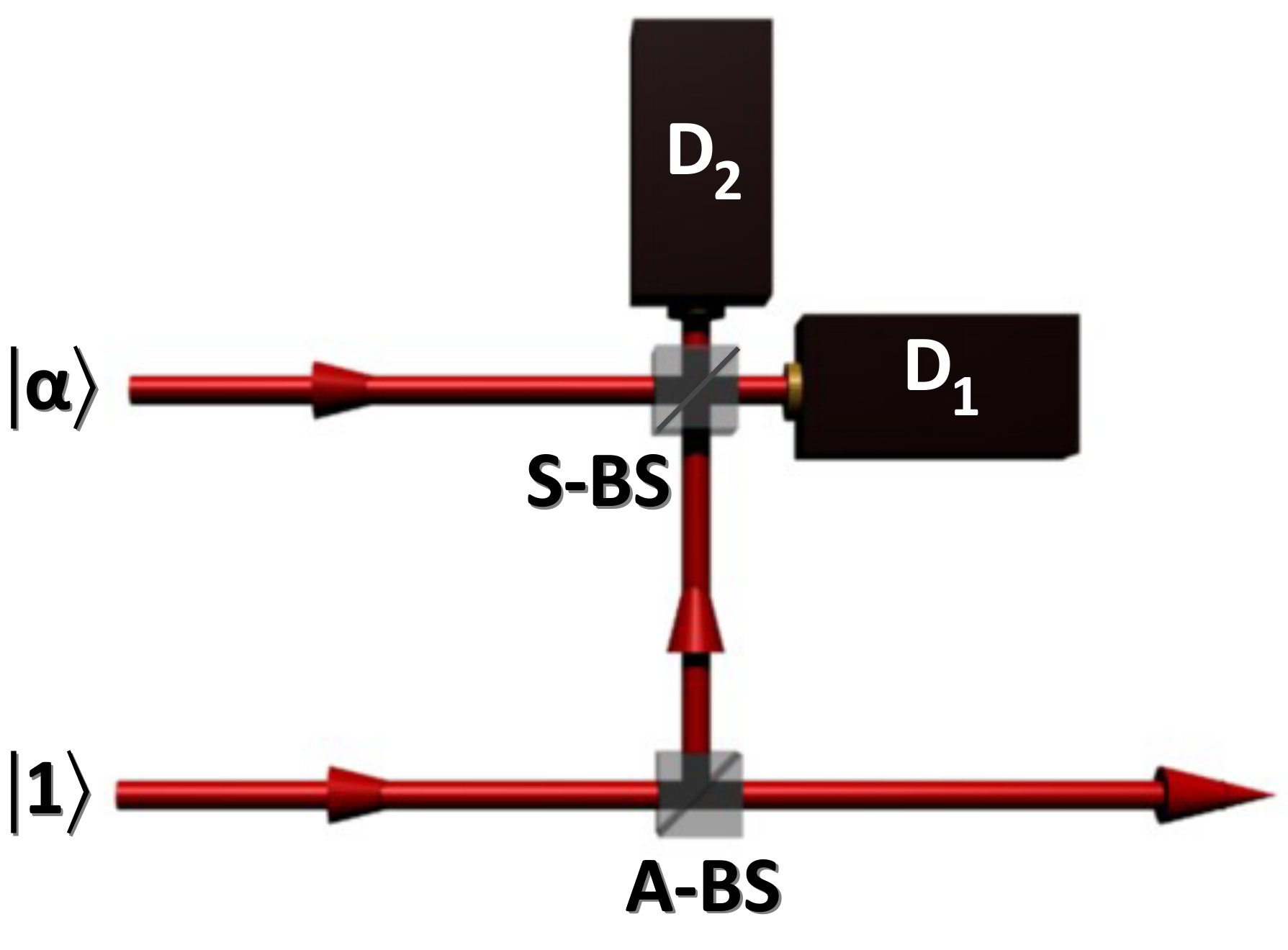}
\caption{Conceptual layout of the noiseless amplifier. A single photon is split on an asymmetric beamsplitter (A-BS). The input state $\ket{\alpha}$  is superposed with reflected output of the A-BS on asymmetric beamsplitter (S-BS). A successful run of the amplifier is flagged by a single photon event on detector $D_1$ and no photons on detector $D_2$. The transmitted mode constitutes the output mode of the amplifier, and is approximately in an amplified state $\ket{g\alpha}$, conditioned on the right detection events, as described by eq. (1).}
\end{figure}


More in detail, a coherent state $\ket{\alpha}$  is fed into the input mode of the amplifier; at the same time an auxiliary single photon beam is provided, and split onto an asymmetric beam splitter (A-BS) with reflectivity $r$; this prepares the two-mode entangled state $\sqrt{1-r^2}\ket{1}_T\ket{0}_R{+}r\ket{0}_T\ket{1}_R$ , where $T\,, R$ denote the output modes of the A-BS. We perform a collective measurement on the input state and part of the entangled state, in our case the $R$ mode; this consists in superposing them on a symmetric beamsplitter (S-BS), and performing photon counting at the outputs. A successful event is flagged by a single photon detection by the detector $D_1$, and no photons detected by the detector $D_2$; conditioned on this event, the (non normalised) state of the $T$ mode, which represents the output of our amplifier, is \cite{ralph09}
\begin{equation}
e^{-\frac{\|\alpha^2\|}{2}}\frac{r}{\sqrt{2}}\left(\ket{0}+\frac{\sqrt{1-r^2}}{r}\alpha\ket{1}\right)
\end{equation}
The output is thus prepared approximately in the coherent state $\ket{g\alpha}$, with $g{=}\frac{\sqrt{1-r^2}}{r}$; the probability of this event is given by the squared norm of the state (1): $P{=} 
e^{-{\|\alpha^2\|}}\frac{r^2}{2}\left(1+g^2\|\alpha^2|\right)$. Events where $D_2$ detects one photon and $D_1$ detects none can still be accepted by using an active phase modulation \cite{ralph09}. The main limitation of the amplifier is the size of input state: for its correct operation it is necessary that $g^2\|\alpha^2\|\ll1$. Larger coherent states can be amplified by splitting the input into several modes, each one with an acceptable size for the amplifier. These modes are then amplified individually, and finally recombined non-deterministically on a single mode \cite{ralph09}. Here we will focus on small values of $\|\alpha^2\|$, which are relevant for continuous-variable quantum cryptography \cite{leverrier09}, and show explicitly how the gain is degraded when $\|\alpha^2\|$  becomes too large.

Single photons are produced by using spontaneous parametric down-conversion in a non-linear crystal. This process generates photon pairs in two correlated modes; the presence of a single photon on one mode is inferred by a click on a single photon detector $D_0$ placed on the other ÒtwinÓ mode \cite{alexei06}. Our down-conversion source is based on a 100$\mu$m thick KNbO$_3$ slab, pumped by doubled Ti:Sa laser pulses ($P\sub{max}{=}3.3$mW, $\lambda_p{=}423.5$nm, $\Delta t{=}220$fs, repetition frequency $\Delta \nu{=}800$kHz). Phase-matching is temperature-tuned to obtain frequency degenerate emission at an angle $\sim3^\circ$. The amplifier works conditionally on a coincidence count between $D_0$ and $D_1$. Due to the limited efficiency of our single-photon detection, $D_2$ can be dropped from the actual implementation without significantly affecting the performance of the amplifier. 

We used homodyne detection and a maximum-likelihood reconstruction algorithm \cite{lvovsky04} to determine the Wigner quasi-probability distribution of the output of our amplifier for several values of $\|\alpha^2\|$. A nominal value $g{ =}2$ Ð corresponding to a 6dB gain in intensityÐ was set by adjusting the A-BS. Each state tomography is reconstructed from a set of 200,000 points divided into 12 histograms according to the measured quadrature. The measured success rates depend on the amplitude, and ranges from $\sim1\%$ for $\|\alpha\| \simeq 0.1$ up to $\sim6\%$ for $\|\alpha\|\simeq1$. The Wigner functions shown in Fig. 2 summarize the behaviour of the amplifier for growing input amplitudes: even for small amplitudes, $\|\alpha\|{=}0.1$, one can observe small departures from the circular shape of a coherent state, in particular different widths along the amplitude quadrature $X$ and the phase quadrature $P$. As the amplitude grows, $\|\alpha\|{=}0.25$, and $\|\alpha\|{=}0.5$, those departures become more important, and the non-gaussian character of the output state in eq.(1) clearly appears.


\begin{figure*}
\includegraphics[viewport=110 40 1000 650, clip, width=1.8\columnwidth]{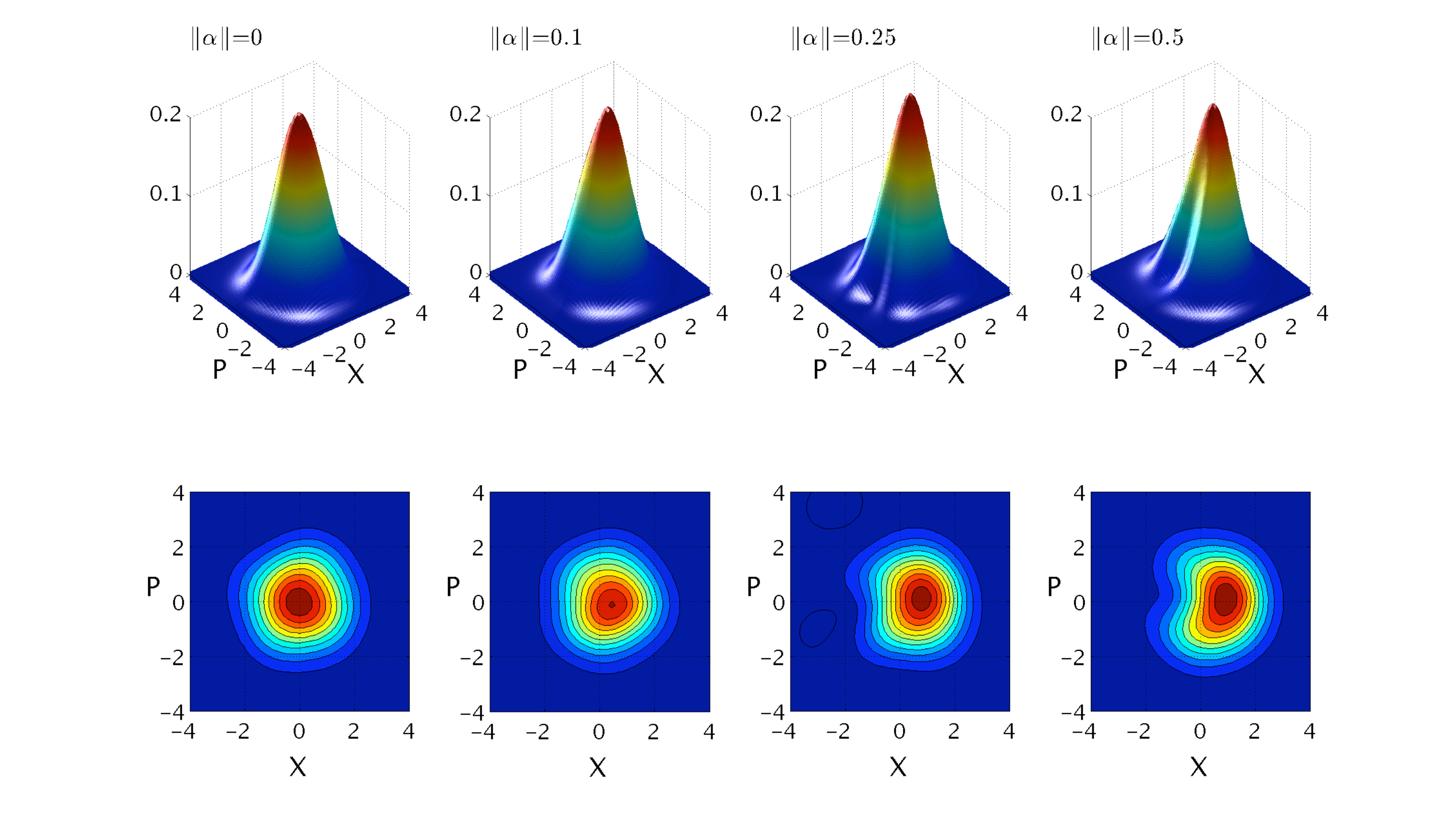}
\caption{Experimental results for the Wigner functions illustrating the evolution of the output state. For each value of  $\|\alpha\|$ we show a 3D and a contour plot of the Wigner function. As the input amplitude   grows, non-trivial structures appear in the state, due to its non-Gaussian character (a superposition of vacuum and one-photon states, eq.(1) ). These results are obtained directly from raw homodyne data, without corrections for the detection efficiency. The value of $\alpha$ is arbitrarily chosen to be real and positive, but the results would be the same for any other choice, since the amplifier gain is phase - independent.}
\end{figure*}

We quantify the effective amplification by introducing an effective gain:
\begin{equation}
g\sub{eff}=\frac{\langle X\sub{out}\rangle}{\langle X\sub{in}\rangle}
\end{equation}
where $X\sub{out}$, $(X\sub{in})$ is the amplitude quadrature of the output (input) field. 

We have used two different methods for characterising the input and output states : the output is analysed with the homodyne detection, while the input is measured using a photon counting avalanche photodiode. This allows us to characterize both beams while the amplification scheme is running. On the input side, the amplitude $\|\alpha\|$ is measured with the detector $D_1$, relating the observed count rate $C$ when blocking the single photon beam; calling $\mu$ the detection efficiency, the value of $\|\alpha\|$  is calculated from the relation\cite{wall08} $C{=}\Delta\nu\left(1-e^{-\mu\|\alpha^2\|}\right)$  . From this measurement, we can obtain the value $\langle X\sub{in}\rangle{=}2\alpha$, for $\alpha$ real. This evaluation has been checked to be fully consistent with the homodyne result, directly performed on the input beam when calibrating the system. 

We consider the values $\langle X\sub{out}\rangle$ and $\langle X\sub{in}\rangle$  just before and just after the amplifier. The ratio of these two quantities would be unchanged if we use the values measured with the same homodyne efficiency $\eta\sub{HD}$, since $\langle X\sub{m}\rangle{=}\sqrt{\eta\sub{HD}}\langle X\rangle$. The homodyne efficiency is estimated as $\eta\sub{HD}{=}0.68$, and originates from imperfect mode-matching (0.9), limited optical transmittivity (0.87), and limited quantum yield of the photodiodes (0.97). Experimental data are compared with a model taking into account the main imperfections of our setup: limited quality of our single photon state, due to multi pair emission and parasite processes\cite{alexei06,alexei07}; imperfect mode-matching between the single photon and the coherent beams; finite photon counting detection efficiency. Our entangled resource is still satisfactory for small coherent states ($\|\alpha\|<0.1)$ , for which the observed gain remains close to the target value $g{=}2$ (Fig.3a). 

The noiseless behaviour of our amplifier is analysed in terms of its ``equivalent input noise" (EIN), 
also called ``noise referred to the input"  \cite{grangier92,roch93, poizat94, grosshans03}:
\begin{equation}
N\sub{eq}{=}\frac{\langle \delta X^2\sub{out}\rangle}{g\sub{eff}^2}-\langle\delta X^2\sub{in}\rangle
\end{equation}
where $\langle \delta X^2 \rangle$ is the variance of the $X$ quadrature just at the output of the amplifier, and is related to the measured value $\langle \delta X^2\sub{m}\rangle$ by the relation $\langle \delta X^2\rangle{=}1+\frac{\langle \delta X^2\sub{m}\rangle-1}{\eta\sub{HD}}$. This figure is the quantum optical analogue to the one adopted in electronics \cite{roch93, poizat94}; it tells how much noise must be added to the input noise level, in order to ÒmimicÓ the observed output noise for the given gain. In Fig. 3b, $N\sub{eq}$ is shown as a function of the input amplitude; we report both the minimal and the maximal EIN, corresponding to the $X$ and $P$ quadratures respectively (Fig. 2), and the EIN averaged over the 12 quadratures corresponding to our histograms. We also report the predicted $N\sub{eq}$ obtained with a phase-independent parametric amplifier driven at gain $g\sub{eff}$. Our data demonstrate how the noiseless amplification actually occurs for all quadratures at the same time; the amplified state remains approximately round when $\|\alpha\|{<}0.1$. 

\begin{figure*}
\includegraphics[viewport=80 80 1050 520, clip, width=1.5\columnwidth]{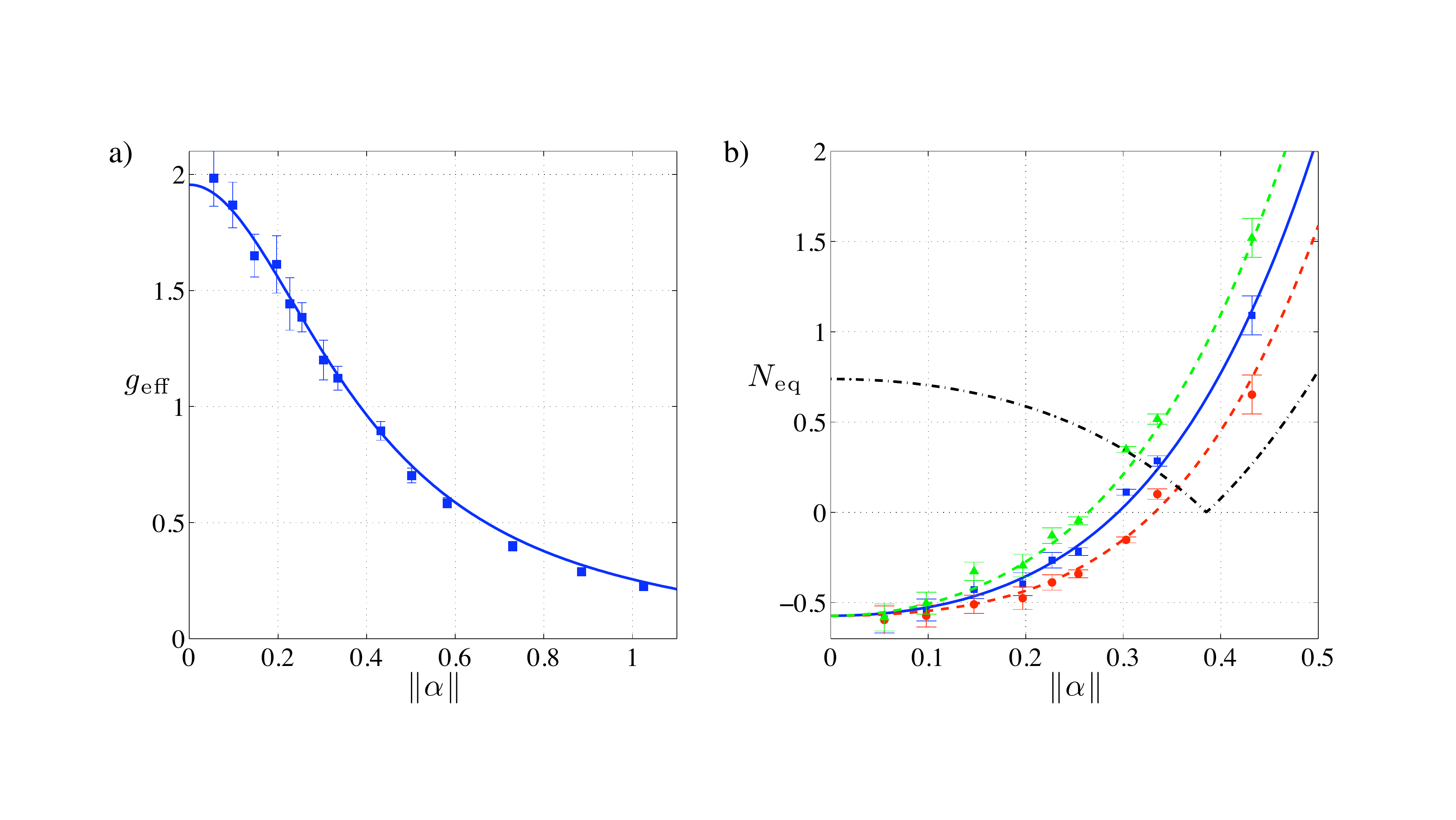}
\caption{Experimental test of the noiseless optical amplifier. a): effective phase-independent gain as a function of the input state amplitude  . The solid line is the prediction from a model accounting for the main imperfection of our setup. The relevant values of $\|\alpha\|$  are smaller than 0.1. b): since the noise of the amplified state is not fully circular in phase space (see Fig. 2), homodyne detections with different phases will see different noises. Therefore we plot the average EIN ($\blacksquare$), maximal EIN ($\blacktriangle$), and minimal EIN ($\bullet$) as a function of the input state amplitude $\|\alpha\|$ . Our model predicts the solid line for the average EIN, and dashed lines for the minimal and maximal EIN in the output state. We also report as a reference (dotted line) the minimal noise attained with a non-degenerate parametric amplifier \cite{caves82} for $g\sub{eff}>1$  and with a beamsplitter model \cite{wall08} for $g\sub{eff}<1$.  For small $\|\alpha\|$ , $N\sub{eq}$ is clearly negative. }
\end{figure*}

Some excess noise is present mostly due to multi-photon events on the auxiliary mode. This noise can be reduced only at expenses of the single photon generation rate. The reported value represents the best experimental trade-off between count rate and excess noise we have achieved on our setup. 

The EIN parameter is always positive in ordinary amplifiers, as these cannot improve the quality of the entering signal. In the present case, it may become negative for specific ÒheraldedÓ events; obviously, when considering the whole set of events, we always observe a behaviour compatible with quantum mechanics. Indeed, we can give a simple argument to show that the amplifier cannot increase the overall information if used at the receiving site of a transmission line. Let us consider the mutual information $I\sub{AB}$ between two parties sharing a Gaussian distribution of coherent states \cite{grosshans03}: $I\sub{AB}{=}\frac{1}{2}\ln(1+\textsc{r})$, where $\textsc{r}$ denotes the signal-to-noise ratio. The amplifier modifies the expression above as: $I\sub{AB}^{\rm amp}{\leq} P  \ln(1+g^2\textsc{r})$, where $2P$ is the success probability when allowing for both heralding events. In the limit of small coherent states one gets: $$I\sub{AB}{\leq}\frac{r^2}{2}g^2\textsc{r}{=}\frac{1-r^2}{2}\textsc{r}\simeq(1-r^2)I\sub{AB}.$$ 
This shows that the success probability of amplification is small enough not to increase the overall mutual information, remaining thus consistent with the general limits imposed by quantum mechanics.

Our investigation demonstrates that some processes that are forbidden with unitary operations can be actually observed in experiments based on quantum measurement and post-selection. Furthermore, we have shown how our amplifier is quite robust against many experimental imperfections, making it valuable resource for quantum communication. This opens the way to the application of our device to non-deterministic entanglement distillation protocols.

\emph{Acknowledgements}  This work is supported by the EU project COMPAS and the ANR SEQURE. F.F. is supported by C'Nano -  \^Ile de France. M.B. is supported by the project `MCQM' of the RTRA `Triangle de la Physique'.

\end{document}